# GENETIC CODE AS A HARMONIC SYSTEM

Miloje M. Rakočević

*Department of Chemistry, Faculty of Science, University of Niš, Ćirila i Metodija 2, Serbia* *

**Abstract**.

In a certain way, this paper presents the continuation of the previous one which discussed the harmonic structure of the genetic code (Rakočević, 2004). Several new harmonic structures presented in this paper, through specific unity and coherence, together with the previously presented (Rakočević, 2004), show that it makes sense to understand genetic code as a set of several different harmonic structures. Thereby, the harmonicity itself represents a specific unity and coherence of physico-chemical properties of amino acid molecules and the number of atoms and/or nucleons in the molecules themselves (in the form of typical balances). A specific Gauss' arithmetical algorithm has the central position among all these structures and it corresponds to the patterns of the number of atoms within the side chains of amino acid molecules in the following sense: G+V = **11**; P+I = **21**; S+T+L+A+G = **31**; D+E+M+C+P = **41**; K+R+Q+N+V = **61**; F+Y+W+H+I = **71**; (L+M+Q+W) + (A+C+N+H) = **81**; (S+D+K+F) + (T+E+R+Y) = **91**; (F+L+M+S+P) = (T+A+Y+H+I) = (Q+N+K+D+V) = (E+C+W+R+G) = **51**. Bearing in mind all these regularities it make sense to talk about genetic code as a harmonic system. On the other hand, such an order provides new evidence supporting the hypothesis established in the previous paper (Rakočević, 2004) that genetic code has been complete from the very beginning and as such was the condition for the otigin and evolution of life.

## 1 INTRODUCTION

In the previous paper we have presented a strictly harmonic structure of the genetic code (Table 1 in Rakočević, 2004), consisting of 4 x 5 canonical amino acids (AAs), which follows from two arrangements of amino acid pairs; the first arrangement as presented in Table 1 and the second one as presented in Table 2. In this paper however we will present several other harmonic structures such that they altogether show that the genetic code is a kind of harmonic system. Thereby, the harmonicity itself represents a specific unity and coherence of physico-chemical properties of amino acid (AA) molecules and of the number of atoms and/or nucleons in them (in the form of typical balances).

* E-mail: m.m.r@eunet.yu

The starting point in this new analysis of harmonicity of genetic code is the system of 16 AAs of alanine stereochemical type (Survey 1.1 in Rakočević and Jokić, 1996 and Table 1 in this paper). That system presents the relations between amino acid *pairs* through a natural (chemical) distinction into 1-2-3-2 pairs, with the balance of the number of atoms (86:86) along two zigzag lines. [The system of 1-2-3-2 pairs: aliphatic AAs A & L, chalcogene S-T & C-M, basic and acidic plus acid (amide) derivatives K-R, D-E & N-Q, and aromatic AAs: H-W & F-Y; Zigzag line actually represents the first possible periodicity within a „periodic system“ consisting of two columns („groups“) and several rows („periods“).]

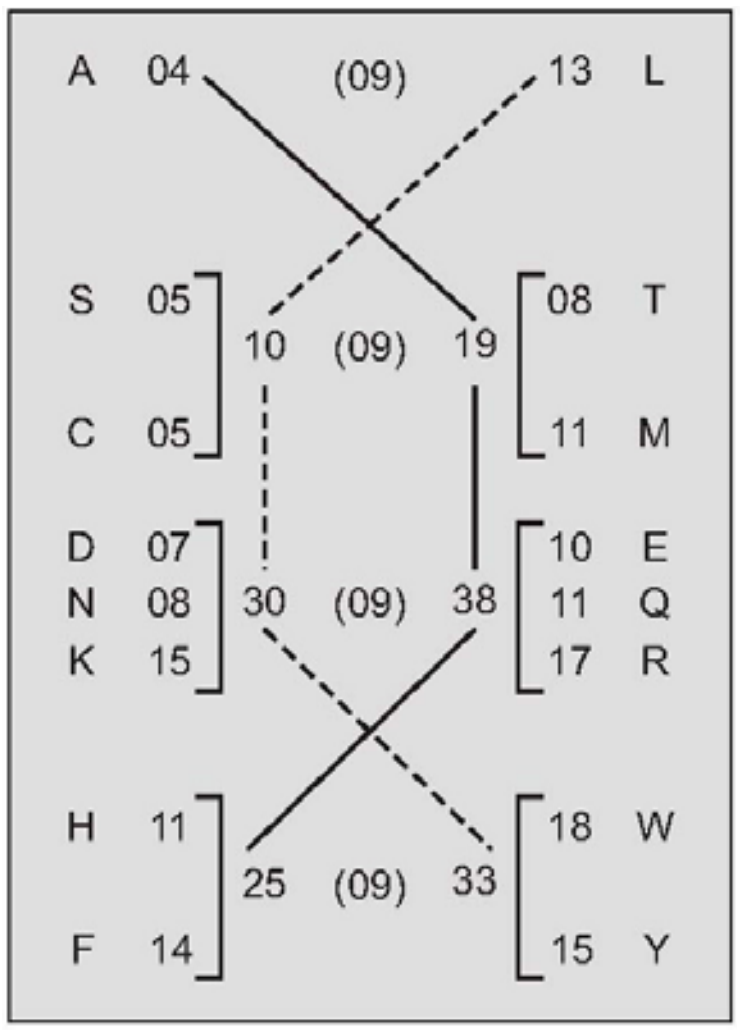

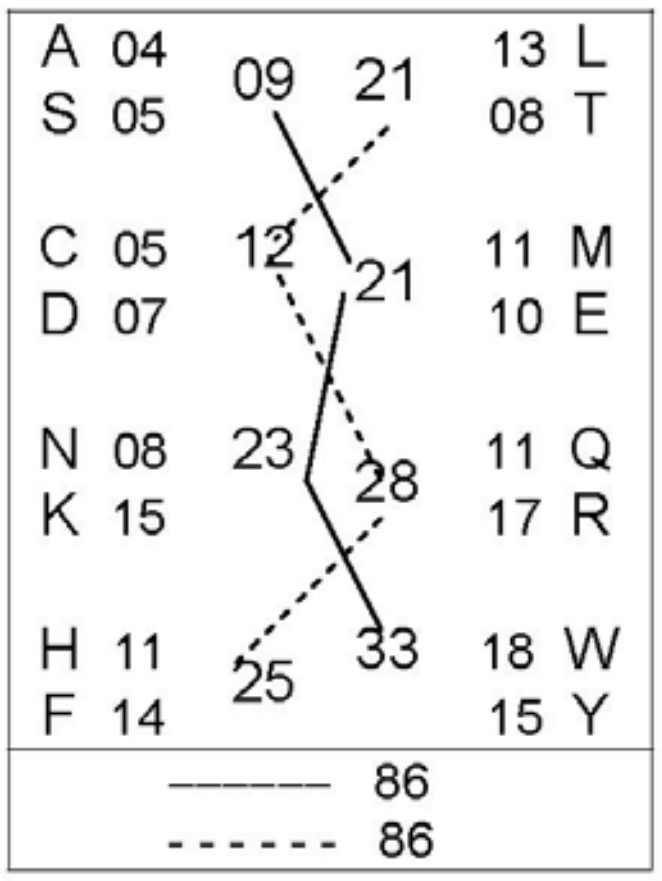

**Table 1 (left).** The 16 amino acids of alaninic stereochemical type (I). This Table is presented in our previous paper (Survey 1.1 in Rakočević & Jokić, 1996). The 8 amino acid pairs belong to the alaninic stereochemical type (Popov, 1989), that is to the class of „non-contact“ amino acids. (About „contact“ and „non-contact“ amino acids *see* in Remark 1.) Hierarchy observes chemical classification into aliphatic (as the first) and aromatic (as the second) amino acids, whereby the order of pairs has been determined by the mass of the first member in the pair.
**Table 1.1 (right).** The 16 amino acids of alaninic stereochemical type (II). This Table originates from Table 1, in the way as explained in the text.

The order of the pairs itself has been determined so that the subclass of aliphatic AAs comes first, followed by the subclass of aromatic AAs. Within each subclass the order is dictated by the number of atoms in the side chain of each first member of an amino acid pair.

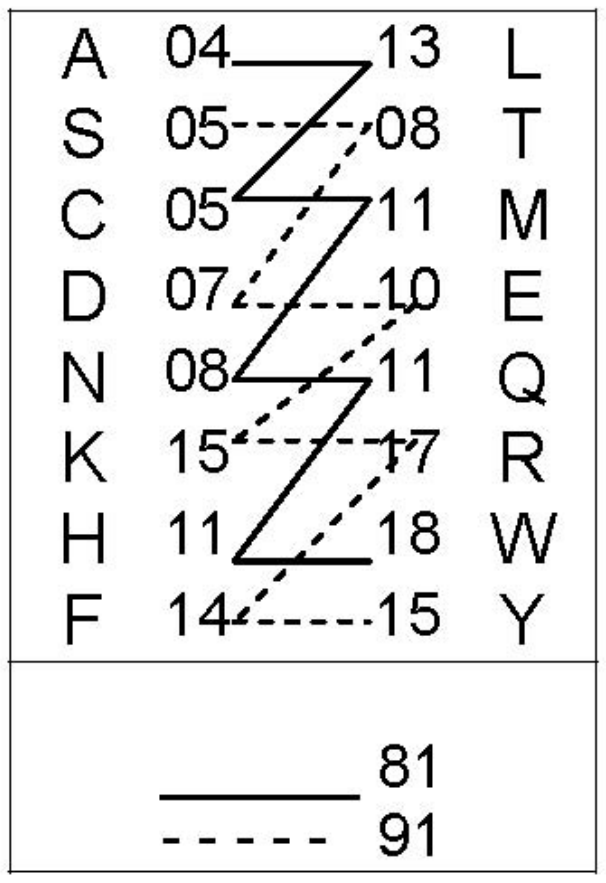

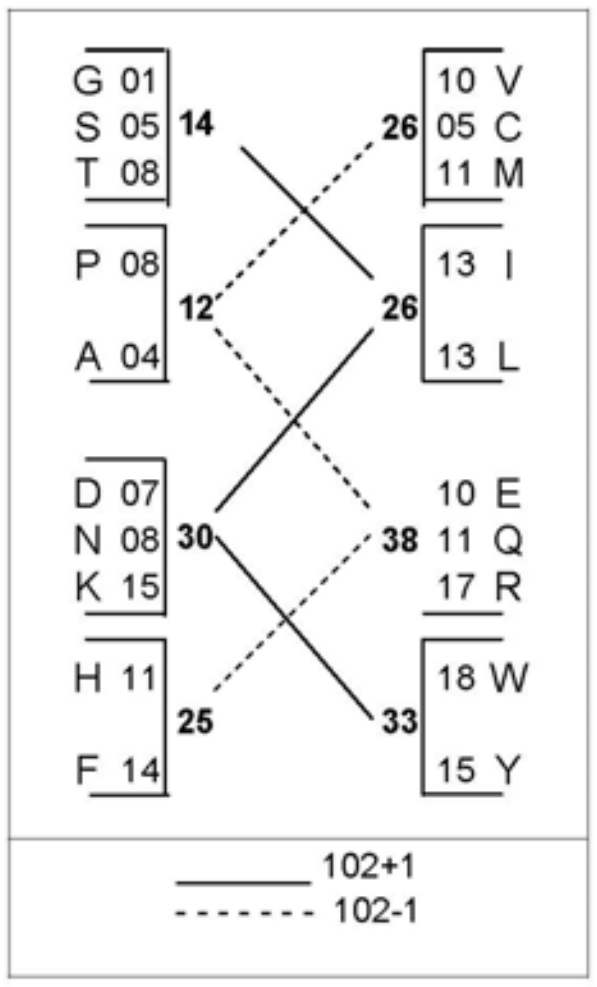

**Table 1.2 (left).** The 16 amino acids of alaninic stereochemical type (III). This Table originates from Table 1.1. On two zigzag lines here we have the connection between amino acid singlets whereas in Table 1.1 between the doublets.
**Table 2 (right).** The atom number balance directed by two classes of enzymes aminoacyl-tRNA synthetases (I). This arrangement of amino acid pairs follows from the first one, presented in Table 1 after hard regularities, given in a previous paper (Rakočević, 1998b). On the first (full) zigzag line, there is 102+1 whereas on the second (dotted) line 102-1 atoms. Arithmetic mean for both: 102±1. Class I of enzymes aminoacyl-tRNA synthetases handles the larger amino acid molecules (on the right) whereas the smaller molecules (on the left) are handled by class II (after Rakočević, 1998b).

## 2 A NEW ARRANGEMENT OF PAIRS

The first question (in this analysis) is the issue of possible rearrangement of the system of eight molecule pairs of AAs of alanine stereochemical type (Table 1), so that the formally unbalanced arrangement of molecule pairs (1-2-3-2), reflecting a possible chemical classification, is transformed into a balanced one (2-2-2-2), maintaining the balance expressed also in the number of atoms along two zigzag lines, regardless of the fact that then the initial classification will be „disrupted". An affirmative answer to this question (repeatedly obtained ratio 86:86 in the atom number along two zigzag lines) is presented in Table 1.1.

Where then did such result come from? The only acceptable explanation is that the said „disruption" of classification is still only an illusion. Once established hierarchy of pairs, through strictly determinated neighbourhood, both chemically (physico-chemical charactiristics) and formally (atom number) must possess at least a unit degree of freedom, so new associations/dissociations of molecules, i.e. molecule pairs, are possible. Thus, from the chemical point of veiw, the said associations/dissociations in a new rearrangement look like this: A-L & S-T, D-E & C-M, K-R & N-Q. By this, in the subclass of aromatic AAs the order is not changed, in the transition from the arrangement given in Table 1 to the one given in Table 1.1.

Indeed, it makes sense, i.e. chemical justification, for *the first* – the oxygen substituted derivatives (S-T) to go together with the original aliphatic molecules (A-L); and also for *the second* – the oxygen substituted compounds to go together with *the first* – the oxygen analogue molecules: D-E with C-M. Finally, it is entirely logical that nitrogen AAs are also found together: K-R with N-Q. [If S-T is the first substituted pair, then D-E is the second; besides, the oxygen pair S-T and sulphuric C-M are analogue in a class of chalcogene AAs.]

## 3 A GAUSS' ARITHMETICAL ALGORITHM

The presence of the sulfur AAs in a set of 20 canonical AAs represents a "nonserial" substitution in the sense that these are the only two AAs whose functional group owns a chemical element located lower than the second period of Periodic system of chemical elements, in the third period more precisely. The question is also raised here why the sulphur was "selected" and not phosphorus, for example? Why (–SH), and not ($-PH_2$) functional group? The answer is that 2 atoms (–SH) fit together, while 3 atoms ($-PH_2$) do not fit into a very specific *arithmetical algorithm*, which we will designate as "Gauss' algorithm"[1] (for reasons in footnote 1 stated) (Figures 1-2 in relation to Tables 3-4**)**.

[1] There are several anegdotes about Gauss as a boy and his understanding of mathematics at his earliest age. One of them say that the teacher gave an exercize to the class of nine year olds in which Gauss was a pupil, to add all numbers from 1 to 100, thinking he will have enough free time to do his "personal things", but also giving the opportunity to potential geniuses... Indeed, less than three minutes passed when little Gauss came with the result: 5050. Asked how calculated it, he explained that he added the first and the last number (1+100 = 101), then the following and the first preceding one (2+99 = 101) and so on. Since there are 50 such pairs, he multiplied 50 x 101 and obtained the requested result. Now, if the teacher had given the exercize to add all numbers not from 1 to 100, but from 1 to 101, and that the results are distributed in 10 columns, Gauss would get the results as in Tab. 3, whose first order presents, as we can see, the "patterns" of the number of atoms in the rows and columns of the system "5 x 4" of AAs in genetic code (Figure 1).

| | (a) | | | | |
|---|---|---|---|---|---|
| S05 | T08 | L13 | A04 | G01 | 31 |
| D07 | E10 | M11 | C05 | P08 | 41 |
| K15 | R17 | Q11 | N08 | V10 | 61 |
| F14 | Y15 | W18 | H11 | I13 | 71 |
| 91 | | 81 | | | |

| | 91 | (b) | 81 | |
|---|---|---|---|---|
| 29 | S05 | T08 | M11 | C05 |
| 36 | D07 | E10 | Q11 | N08 |
| 49 | K15 | R17 | L13 | A04 |
| 58 | F14 | Y15 | W18 | H11 |
| 32 | G01 | V10 | I13 | P08 |
| | 11 | | 21 | |

| | |
|---|---|
| 2 | 20-7 |
| 3 | 20+7 |
| 4 | 46-5 |
| 5 | 46+5 |
| 6 | **86-1** |
| 7 | **86+1** |
| 8 | 140+5 |
| 9 | 140-5 |

**Figure 1 (left).** The distribution of amino acids according to Gauss' algorithm; (a) the distribution of amino acids has been derived from Table 1.1, by presenting two pairs here in one row as the "pair of pairs". In the beginning of each row obtained in this way one "contact" amino acid has been associated, by the increasing molecule mass (about "contact" and "non-contact" amino acids see in Remark 1). Atom number (in amino acid side chains) in the rows and columns generated in this way corresponds, one hundred per cent, to the quantums from the first row of Gauss' algorithm of adding numbers from 1 to 101 (Table 3). Dark tones: Class I of amino acids handled by class I of enzymes aminoacyl-tRNA synthetases; light tones: Class II. Obviously, chemically related groups of AAs have been "taken off" by 0, 1 and 2 steps, respectively. By zero "steps" in aromatic; by one step in chalcogenic AAs (M, C in relation to S, T) and carboxylic (carboxylic AAs D & E in relation with their amides N & Q); by two steps in source aliphatic AAs: A, L & K, R. (b) The distribution is the same as in (a) but "leveling" has been conducted here according to the chemical properties of molecules, so there is no more taking off. The class of contact AAs has been added to the beginning of columns, instead of to the beginning of rows. Shading is the same as in (a). Chemical "leveling" conducted in this way is correspondent with the determination of structure with a specific algorithm – algorithm of symmetry through cyclization (Figure 1.1).

**Figure 1.1 (right).** Arithmetical algorithm of symmetry through cyclization, correspondent to amino acid splitting into four chemical classes, as it is shown in Figure 1, on the right. Within two inner classes (AA side chains) there is 81-1 whereas within two outer classes 81+1 of atoms. The steps of the algorithm: 1. Choose two adjacent numbers from decimal scale; 2. Take the squares of both; 3. Move one modular cycle more (in module 9), for example from 49 to 58; 4. Go back for one half, for example from 58 to 29. From the Figure it is clear that only one solution is possible (balance in accordance to principle of minimum change).

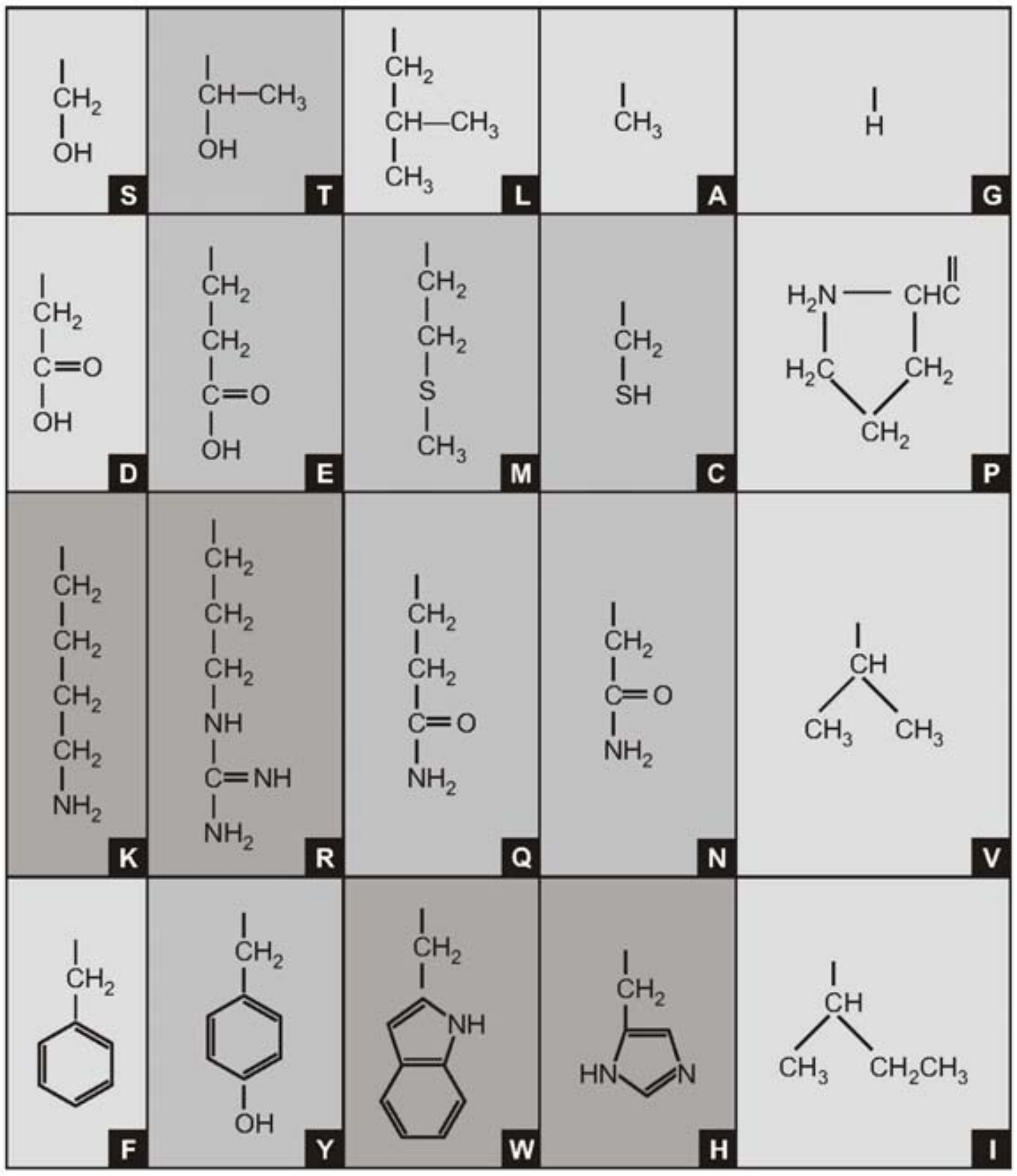

**Figure 1.2**. The structure of amino acid molecules. The simplest amino acid is glycine (G) whose side chain is only one atom of hydrogen. It is followed by alanine (A) whose side chain is only one $CH_3$ group, which is the smallest hydrocarbon group. There are total of 16 amino acids of alaninic stereochemical type (“non-contact” amino acids) with one $CH_2$ group each between the “body” and the “head”. The glycinic type contains glycine (G) only; valinic type contains valine and isoleucine (V, I); The last stereochemical type is prolinic with proline (P) which represents the inversion of valine in the sense that the “triangle” of three $CH_2$ groups for the “head” is not bound by the basis, therefore not only with one but with two $CH_2$ groups (Popov, 1989; Rakočević & Jokić, 1996). Light tones (G, P, V, I & A, L, S, D, F): invariant AAs; most dark tones (K, R, W, H): most variant AAs; less dark tones (T, E, Q, M, N, C): less variant AAs. (Cf. Section 4.1.)

| | | | | | |
|---|---|---|---|---|---|
| T 08 | M 11 | C 05 | **<u>S</u> 05** | G 01 | **31 - 1** |
| **<u>D</u> 07** | E 10 | L 13 | A 04 | P 08 | **41 +1** |
| **<u>K</u> 15** | R 17 | Q 11 | N 08 | V 10 | **61 ± 0** |
| **<u>F</u> 14** | Y 15 | W 18 | H 11 | I 13 | **71 ± 0** |

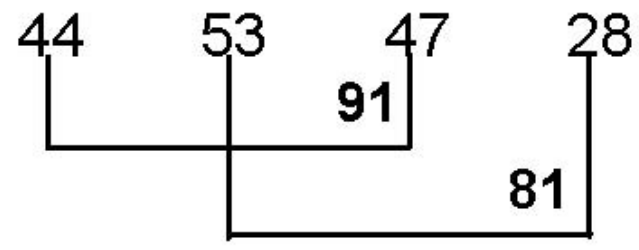

**Figure 2**. The distribution of amino acids according to Gauss' algorithm, with a minimal change: for ±1 atom in first two rows and for one "take of" in columns. The distribution of amino acids has been derived from Table 2.1 in the same manner as Figure 1 from Table 1.1. Dark and light tones show the changes within the AAs columns going from Figure 1 to Figure 2 (for example: T,E,R,Y in column in Figure 1).

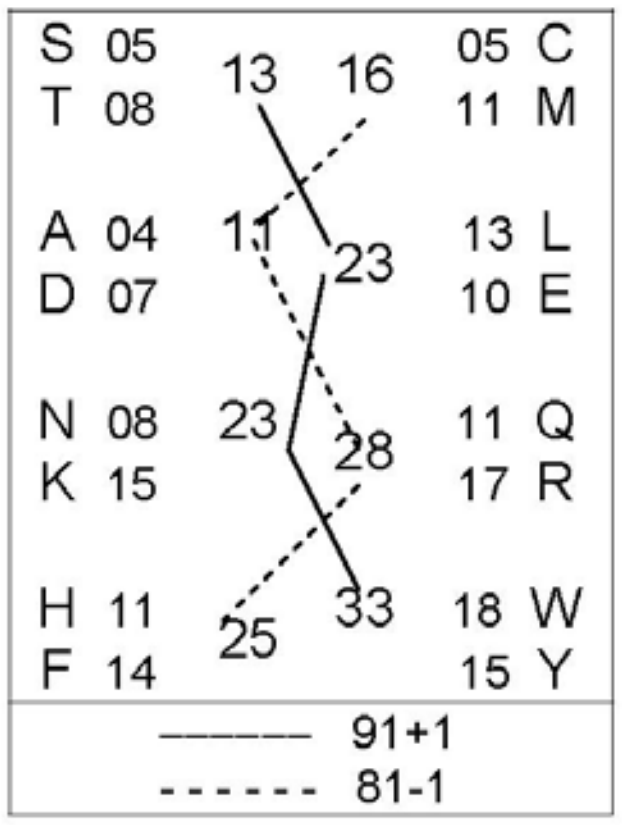

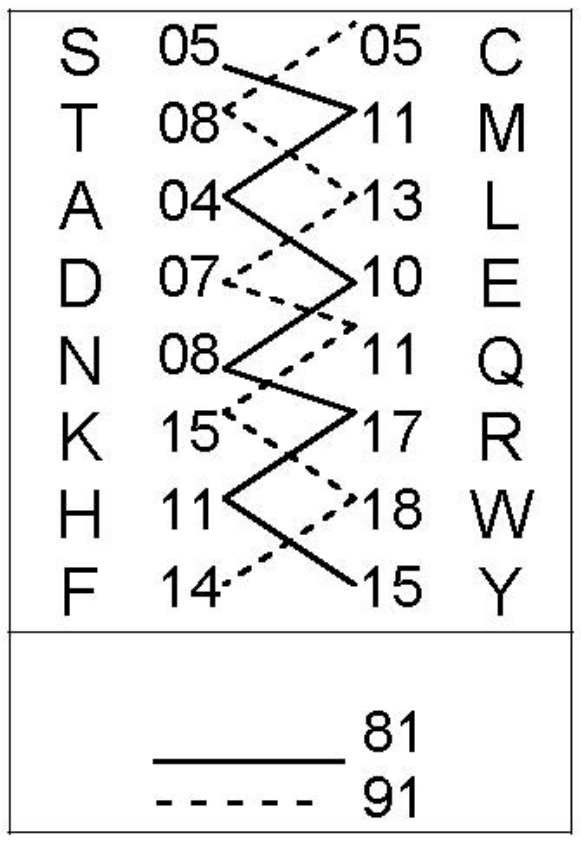

**Table 2.1 (left).** The atom number balance directed by two classes of enzymes aminoacyl-tRNA synthetases (II). All is the same as in Table 2, except the fact that contact AAs (G-V and P-I) are excluded. The arrangement itself is analog to this one in Table 1.1, the AAs are splitting into 4 x 4 sets.

**Table 2.2 (right).** The atom number balance directed by two classes of enzymes aminoacyl-tRNA synthetases (III). This Table follows from Table 2.1 in the same manner as Table 1.2 from Table 1.1.

However, apart from the atom number, the type of functional group (actually the unity of these two factors) determines priority in the „selection" of one or the other amino acid, in the following sense. If we can consider hydrocarbon molecules as main organic compounds, which is a fact actually, then (–OH) derivatives, with 2 atoms in a functional group, are the first possible ones in the act of „copying“ the head of an amino acid to the body (in which copying the principle of *self-similarity* is realized, moving from a part of a molecule to another); only then does the copying of (–$NH_2$) group with 3 atoms follow and finally, the copying of (–COOH) group with 4 atoms, which represents the highest degree of oxydation (and substitution) in the main aliphatic chain. All three copyings are related to functional groups with a single covalent bond, while the copying of (=CO) group which possesses double bond does not occur. However, it is not „neglected“, since this group is the one through which connection with the amino acid precursors is made. [The amino acid precursor pyruvate, a derivative of acetic acid, possessing a (=CO) functional group, is the starting and central precursor of an amino acid (Rakočević, 1998a).]

The finding that the atom number in side chains of 20 canonical AAs, when they are organized in a strict system, the *pairs of pairs* system, is one hundred percent determined by Gauss' algorithm[2], is demonstrated with the main argument which supports not only the *hypothesis* on the complete genetic code, but also the *thesis* that GC is not of a random but strictly determined nature. After such finding it even makes sense to raise a new hypothesis, for further researches, that this and such genetic code represents a general pattern of the *life code*, valid for the entire universe, wherever the existence of water in its liquid aggregate state is possible, with other corresponding conditions.

## 3.1 Further determinations

The system of eight pairs of AAs of alanine stereochemical type (Table 1) should also be tested from the aspect of possible maintenance of balance of the number of atoms when this system – the „non-contact“ amino acids class – is associated with the remaining three stereochemical types, with the „contact“ AAs class. [*Remark 1*: „Contact“ AAs are those AAs in which there is a direct contact of the „body“ and the „head“, i.e. the side chain with the amino acid functional group, while „non-contact“ AAs are the ones in which contact is mediated by a $CH_2$ group (only with threonine that group is metyl-substituted: $H - C - CH_3$). Contact AAs include three stereochemical types: glycinic (with glycine), prolinic (with proline) and valinic (with valine and isoleucine), while the remaining 16 AAs belong to non-contact AAs, all of alaninic sterechemical type. (For details about stereochemical types, *see* Popov, 1989; Rakočević & Jokić, 1996).]

---

[2] The rules of Gauss' algorithm [algorithm of calculating the sum of all numbers in a row in the sequnce of natural numbers, starting from 1 to a number *n* (1 + 2 + 3 + ... + n)] are obvious: I. *Pairs* of numbers (1, n), (2, n-1), (3, n-2), etc. are formed, for which for the even *n* there are n/2, and for the odd *n* there are (n-1)/2 (Tab. 3), each pair with the sum n+1; II. The value of product (n+1)(n/2), i.e. (n+1)(n-1)/2 is calculated, for even, i.e. odd *n*, respectively; III. The central member of the sequence is added to the obtained products, which is 0 for even *n*, and [(n-1)/2] +1 for odd *n*.

An affirmative answer to the requested testing is presented in Figure 2. How did we come to this result? We have demonstrated that the system of AAs pairs, presented in Table 1, can be „opened“ and with strict rules translated into the system – „the pairs of pairs“ (Figure 1). In one of our earlier works (Rakočević, 1998b) we showed that the same system (Table 1), also according to strict rules, can also be incorporated into the system of two AAs classes, handled by class I and class II of enzymes aminoacyl-tRNA synthetases, respectively (in further text: First enzyme handled system, EHS1, Table 2).

Once we know that, it makes sense to „open“ this system as well (by means of excluding contact AAs, G-V and P-I), and to translate it into „the pairs of pairs“ system (Figure 2). As we can see, only with a small degree of freedom (±1), i.e. with two minimal exceptions from the Gauss' algorithm, a whole new (complete) system is created (Second enzyme handled system, EHS2, as presented in Figure 2).

## 3.2 Enzyme determination and polarity

The second enzyme handled system, EHS2 (Figure 2), can be expanded by adding contact AAs not only to rows but also to columns as presented in Figure 2.1. Nevertheless, what is of special importance here is a new atom number balance within the columns (102:102 through multiples of number 6) we also have a visible distinction of AAs from the aspect of polarity/nonpolarity. Thereby, it is evident from Figure 2 that all these balances and distinctions are possible through one exception in EHS2 in comparison with EHS1: the original order F-Y-W-H must be inverted into order F-Y-H-W.

| | | | | | | | | | | | | |
|---|---|---|---|---|---|---|---|---|---|---|---|---|
| T08 | M11 | C05 | S05 | G01 | 31-1 | T045 | M075 | C047 | S031 | G001 | 199 | 444 |
| D07 | E10 | L13 | A04 | P08 | 41+1 | D059 | E073 | L057 | A015 | P041 | 245 | (**544**) |
| K15 | R17 | Q11 | N08 | V10 | 61±0 | K072 | R100 | Q072 | N058 | V043 | 345 | 811 |
| F14 | Y15 | H11 | W18 | I13 | 71±0 | F091 | Y107 | H081 | W130 | I057 | 466 | (**711**) |
| V10 | G01 | P08 | I13 | | | V043 | G001 | P041 | I057 | | | |
| **54** | **54** | **48** | **48** | | | **310** | **356** | **298** | **291** | | | |
| 102 / 102 | | | | | | 666+00 / 666 – 77 | | | | | | |

**Figure 2.1**. This Figure follows from Figure 2 through an adding of contact AAs not only into rows but also into the columns. The details in the text (Section 3.2).

Polar AAs are positioned as a separate entity, in the form of a specific "island" surrounded by non-polar AAs (Figure 2.1). Within the "island" there are all polar AAs, all except serine, which is separate, whereby the existence of a "commodity" with a certain degree in expressing a degree of freedom in polarity as well. Interestingly enough, there are also three ambivalent AAs (glycine, proline and tryptophan)[3] positioned in a snug "string" at the very edge of the system. The distinction is even more complete when semi-ambivalent histidine is added to these three ambivalent AAs. [We can really regard histidine as a „semi-ambivalent" amino acid since it has neither positive nor negative value in cloister energy, but its value is equal to zero (Figure 5 in Swanson, 1984).]

## 3.3 Atom number multiples and bioprecursors

The presented determination by number 6 (Figure 2.1)[4] can be brought in connection with the determination of canonical AAs (on the binary-code tree) by Golden mean (Survey 2.1 in Rakočević, 1998b)[5]; also with the determination through a specific splitting of AAs in relation to their biosynthetic precursors (Figure 3) as well as with the determination of their positions in the Genetic code table (GCT) (Figure 4).

---

[3] Glycine: after hydropathy is polar; after cloister energy and polar requirement is non-polar; Proline: after hydropathy and cloister energy is polar; after polar requirement is non-polar; Tryptophan: after hydropathy and polar requirement is polar; after cloister energy is non-polar. Hydropathy (Kyte & Doolittle, 1982; Doolittle, 1985); cloister energy (Swanson, 1984); polar requirement (Woese et al., 1966; Konopel'chenko and Rumer, 1975). About the pairing process of AAs through Hydropathy and Cloister energy *see* Survey 1 in Rakočević & Jokić, 1996, and about ambivalence of glycine and proline see Section 3.3 in Rakočević, 2004.

[4] As a noteworthy is the fact that number 6 is the first perfect number. A hypothesis that perfect numbers can appear as determinant of genetic code we published ten years ago (Rakočević, 1997, pp. 60-64).

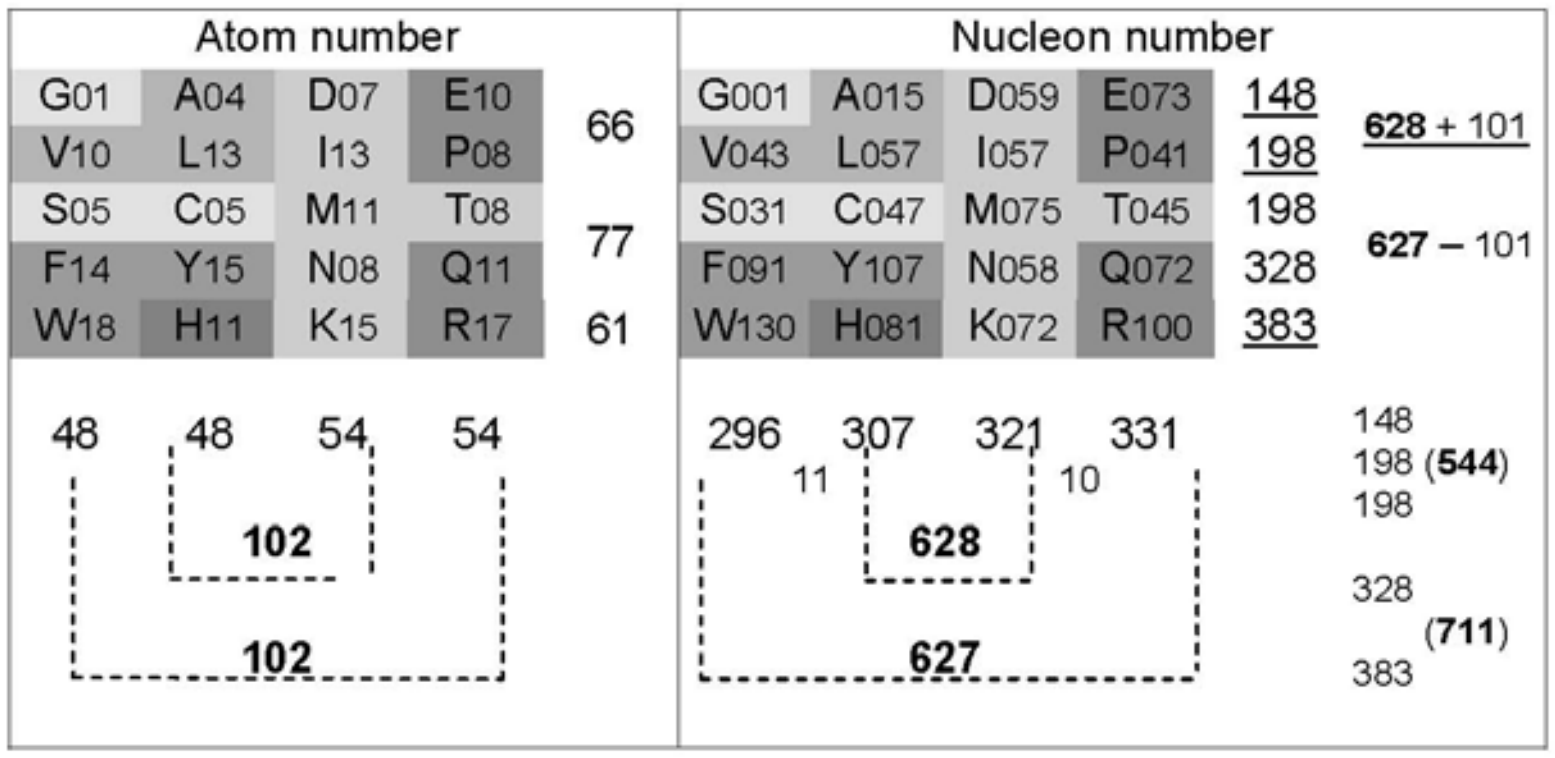

**Figure 3**. The number of atoms and nucleons within side chains of AA molecules in correspondence with the classification of their bioprecursors (*see* text, Section 3.3). On the left side (in both parts of the Figure): very light tones – AAs synthesized through first bioprecursor (3-Phosphoglycerate: G, S, C); middle light tones – AAs synthesized through second bioprecursor (Pyruvate: A, L, V); very dark tones – amino acid synthesized by fifth bioprecursor (Ribose-5-phosphate: H); dark tones – AAs synthesized by sixth bioprecursor (Phosphoenolpyruvate plus eritrose-4-phosphate: F, Y, W). On the right side (in both parts of the Figure): light tones – AAs synthesized by third bioprecursor (Oxaloacetate: T, M, I, D, N, K); dark tones – AAs synthesized by fourth bioprecursor (2-Oxoglutarate, i.e. α-Ketoglutarate: P, E, Q, R).

Figure 3 shows atom and nucleon number balances in correspondence to classification of amino acid bioprecursors into two classes: the first class with four "outer" precursors and the second class with two "inner" precursors. [*Remark 2*: "Outer" bioprecursors are: $1^{st}$, $2^{nd}$, $5^{th}$ and $6^{th}$ and the "inner" ones $3^{rd}$ and $4^{th}$. The order and hierarchy of AAs biosynthetic precursors are given as in our previous work (Table 1 in Rakočević & Jokić, 1996): 1. 3-Phosphoglycerate (G, S, C); 2. Pyruvate (A, L, V); 3. Oxaloacetate (T, M, I, D, N, K); 4. 2-Oxoglutarate, i.e. α-Ketoglutarate (P, E, Q, R); 5. Ribose-5-phosphate (H); 6. Phosphoenolpyruvate plus eritrose-4-phosphate (F, Y, W). Thereby, the first three precursors are *less complex* and last three are *more complex*. At the same time, the first and the two last ones are phospho-precursors, while other three are non-phospho-precursors.]

[5] Rakočević, 1998b, p. 289: „Within seven 'golden' amino acids there are 60 atoms; within their seven complements there are [60+(1 x 6)] and within six non-complements there are {[60+(1 x 6)] + (2 x 6)} of atoms."

| | | | | | | | | | | | |
|---|---|---|---|---|---|---|---|---|---|---|---|
| F14 | T08 | Q11 | E10 | V10 | 51+2 | F14 | S05 | Y15 | C05 | P08 | 47 |
| L13 | A04 | N08 | C05 | I13 | $51-2^3$ | L13 | W18 | H11 | R17 | G01 | 60 |
| M11 | Y15 | K15 | W18 | G01 | $51+3^2$ | Q1 | T0 | N0 | M11 | V10 | 48 |
| S05 | H11 | D07 | R17 | P08 | 51-3 | K15 | A04 | D07 | E10 | I13 | 49 |
| P08 | I13 | V10 | G01 | | | G01 | I13 | P08 | V10 | | |
| 51 | 51 | 51 | 51 | | | *54* | *48* | *49* | *53* | | |
| F091 | T045 | Q072 | E073 | V043 | Out 627+9 | F14 | L13 | S05 | Y15 | G01 | *48* |
| L057 | A015 | N058 | C047 | I057 | | C05 | W18 | H11 | Q11 | P08 | *53* |
| M075 | Y107 | K072 | W130 | G001 | 628-9 In | R17 | M11 | T08 | N08 | V10 | *54* |
| S031 | H081 | D059 | R100 | P041 | | K15 | A04 | D07 | E10 | I13 | *49* |
| P041 | I075 | V043 | G001 | | | I13 | P08 | G01 | V10 | | |
| 295 | 305 | 304 | 351 | | | 64 | 54 | 32 | 54 | 32 | |
| 600 | | 655 | | | | | 86 | | 86 | | |

**Figure 4**. Three "readings" of Genetic Code Table; first on the left side and last two on the right side of Figure. *On the left, up*, the reading through columns: if "contact" AAs (G-V and P-I) are excluded from GCT, the rest of "non-contact" AAs exists in an arrangement of 4 x 4: first quartet (F+L+M+S), second (T+A+Y+H), third (Q+N+K+D+V) and forth (E+C+W+R); if "non-contact" AAs are excluded from GCT, the rest of "contact" AAs exists in an arrangement of 4 x 1: regarding GCT in up-down direction, the first amino acid is P, second I, third V and the fourth G. Altogether: (F+L+M+S+**P**) = (T+A+Y+H+**I**) = (Q+N+K+D+**V**) = (E+C+W+R+**G**) = 51 of atoms each (within side chains); light tones – polar AAs; middle light tones – ambivalent AAs (G, P, W); dark tones – non-polar AAs. *On the left, down*: the order is the same as previous, but with nucleon number balances. *On the righ, up*, the reading through rows – *Py* level. *On the right, down*, the reading through rows – *Py-Pu* level.

## 3.4 Three "readings" of Genetic Code Table

The order of AAs in the very GCT is also determined by multiples of number 6 (Figure 4). We have come to this result with the finding that the order of non-contact AAs should be read after the exclusion of contact ones and vice-versa.

The upper left corner of Figure 4 shows the order of reading by columns, in the classes of four AAs (with association of contact AAs at the beginning of each column and each row, in the adequate sequence). As we can see, atom number in all columns (51 each) is exactly as much as the central point[6] of Gauss' algorithm of „adding numbers to 101“ (Table 3). We can say that these are 102 atoms each in the first and the second half of GCT, where number 102 is also multiple of number 6 (102 = 17 x 6).

| | | | | |
|---|---|---|---|---|
| 01+101 | **11+91** | **21+81** | **31+71** | **41+61** |
| 02+100 | 12+90 | 22+80 | 32+70 | 42+60 |
| 03+099 | 13+89 | 23+79 | 33+69 | 43+59 |
| 04+098 | 14+88 | 24+78 | 34+68 | 44+58 |
| 05+097 | 15+87 | 25+77 | 35+67 | 45+57 |
| 06+096 | 16+86 | 26+76 | 36+66 | 46+56 |
| 07+095 | 17+85 | 27+75 | 37+65 | 47+55 |
| 08+094 | 18+84 | 28+74 | 38+64 | 48+54 |
| 09+093 | 19+83 | 29+73 | 39+63 | 49+53 |
| 10+092 | 20+82 | 30+72 | 40+62 | 50+52 |
| 01+100 | 11+90 | 21+80 | 31+70 | 41+60 |
| 01+101 | **11+91** | **21+81** | **31+71** | **41+61** |
| 01+102 | 11+92 | 21+82 | 31+72 | 41+62 |

| | | | | | | | | | |
|---|---|---|---|---|---|---|---|---|---|
| 0<u>**0**</u>1 $_{1}$ | 01 $_{10}$ | | | | 101 $_{1}$ | 002 $_{10}$ | | | |
| 013 $_{4}$ | 07 $_{10}$ | 06 | 6 | | 101 $_{4}$ | 017 $_{10}$ | 15 | 18 | |
| 025 $_{7}$ | 19 $_{10}$ | 12 | 6 | 1 | 101 $_{7}$ | 050 $_{10}$ | 33 | 18 | 1 |
| 037 $_{10}$ | 37 $_{10}$ | 18 | 6 | 1 | 101 $_{10}$ | 101 $_{10}$ | 51 | 18 | 1 |
| 049 $_{13}$ | 61 $_{10}$ | 24 | 6 | 1 | 101 $_{13}$ | 170 $_{10}$ | 69 | 18 | 1 |
| 05B $_{16}$ | 91 $_{10}$ | 30 | | | 101 $_{16}$ | 257 $_{10}$ | 87 | | |
| ..... | | | | | | | | | |

**Table 3 (left).** The Gauss' algorithm for *n* = 101. The calculating the sum of numbers from 1 to 101 by determining the number of pairs with the same sum. Since here *n* is an odd number, a middle member appears here, which is 51. In the down part of the Table, the first row has been given for the Table with *n* = 100, then the first row for the Table with *n* = 101 and, finally, the first row for the Table with *n* = 102. (The explanation in the text, especially in footnotes 1 and 2.) **Table 4 (right).** The Analogues of number 037 and number 101. Shcherbak (1994; 2003) demonstrated that the number of nucleons in two classes of amino acids (in four-codon and non-four-codon AAs) has been determined with "prime quantum 037", and he also established that out of all double digit numbers written in three positions, number 037 in the decimal numbering system has specific characteristics: all three digits, if they are not reduced to the same digit (as in 3 x 037 = 111), are preserved by multiplying numbers in the limits of module 9 (for example: 1 x 037 = 037; 10 x 037 = 370; 19 x 037 = 703). Shcherbak presented numbering systems with the same characteristics; in other words, he presented analogues of number 037 in these systems, and here they are given on the left side of the Table. With all this, we should note that here determination of continuously generated "steps" is "hidden" here (1 x 1 = <u>1</u>; 1 x 11 = <u>11</u>; 3 x **037** = <u>111</u>; 11 x **101** = <u>1111</u>). With this insight, it becomes clear why number 101 is also specific and why it has been "selected" to be the key determinant of the genetic code.

Nevertheless, there is a determination with the number 6 also by reading the order of AAs by rows, of which (orders) there must inevitably be two: right up and right down in Figure 4. The *right up* arranging of AAs is realized in sach a way that *Py* is read first, followed by the reading of *Pu* (in third codon positions); the *right down* arrangement of AAs represents a „leap“ in the sense that one must first „read“ all AAs in one four-codon family (*Py-Pu* in the third codon position) in one row, then in the next row, etc. Of course, here we also should add contact AAs at the beginning of columns and rows in adequate permutations – adequate from the aspect of achieving balance in the atom number.

[6] We should take note that this is the only point which has been excluded from the numbering system presented in Table 3.

Somewhat unexpectedly, we find the same patterns of atom number, up in the columns and down in the rows: 48-49 and 53-54, which are the same multiples of the atom number we find in the system in Figure 3, with differentiations ±0 and ±1. Moreover, nucleon number in three systems (in Figures 2.1, 3 and 4), also in the relation columns-rows, realizes a specific balance (655 – 544 = 111 and 711- 600 = 111)[7].

Take note that in the system presented in Figure 4, determination with number "51" is shown, not only by columns, but by rows as well, with a typical change, correspondent to module 3/2. [We know that the ratio 3:2 is "the limit of the golden numbers" (Moore, 1994) and a key relation in the triadic Cantor set. (Falconer, 1990, p. XIII: "the middle third Cantor set is one of the ... most easily constructed fractals").]

## 4 LOGIC OF CYCLICITY

The prerequisite and the possbility of unity of physico-chemical characteristics and arithmetical structures, i.e. arithmetical algorithms, are contained in an authentic logic – in *the logic of cyclicity* of the molecules themselves. Namely, by the insight into the validity of Gauss' algorhythm, as a strict determinant of the genetic code, it becomes clear that our earlier presentation (Rakočević & Jokić, 1996) on generating amino acid constituents on the principle of open/closed condition of the side chain now has to be understood as an inevitable *logic of cyclicity*, whereby an open chain presents the zeroth cyclicity.

### 4.1 Variant and invariant Amino acids

In the analysis of these purely chemical relations, we start from Figure 1.2 which graphically presents amino acid molecules according to the same distribution as in Figure 1, which corresponds to Gauss' algorithm. Out of 20 AAs nine appear as *invariant*, in the sense that these are the first possible amino acid molecule-structure solutions; the remaining 11 are *variant*, but the variability is limited by the Gauss' algorithm. Namely, the first row is followed by the second one with only slightly more massive molecules, such that they provide the transition from the quantum-level „31“ to the quantum-level „41“. After the second order significantly more massive molecules must be „selected“ so they could achieve the quantum-level „61“, at the same time „overlapping“ the quantum-level „51“. In the next row the „tempo“ is slower again – again we have a leap of about ten units (to quantum-level „71“). But to make the whole thing more strict and more specific, the columns must be fitted into the algorithm too, two columns into quantum-level „81“ and the remaining two in the quantum-level „91“.

---

[7] Number 111 is the first possible Shcherbak's same-digit pattern, determined by Prime quantum 037 (Shcherbak, 1994) and correspondent to the pattern 1111, determined by Prime quantum 101 (cf. Table 4 where is shown that 3 x 037 = 111 and 11 x 101 = 1111).

The presented double “jump” from quantum-level “41” to quantum-level “61” could explain why, for example, in the third row of the system presented in Figure 1.2, there is a chemically less cognate arginine (with 17 atoms in the side chain) along with lysine, and not the chemically much more cognate ornithine. (Jukes, 1973, p. 24: „I have suggested that arginine displaced ornithine during the evolution of protein synthesis“.) With 12 atoms in the side chain ornithine would find itself in the „baned“ area between two „allowed“ levels. However, bearing in mind the fact that in such a significant biochemical process, such as the biosynthesis of urea, there are interproducts as citrulline, apart from ornithine and arginine, the question is raised why it could not have been the candidate instead of arginine?

In the adequacy of arginine and inadequacy of citrulline lies the entire sense of the logic of choice: apart from formal correspondence (citrulline also has exactly 17 atoms in the side chain, just like arginine) the correspondence in chemical characteristics is also required, as much as possible. So, arginine, with one amino and one *imino* group, is adequate for to make a pair with lysine, while citrulline with one amino and one *carbonyl group* is not. The choice of amino acid molecule with any other number of atoms in the side chain except 17 (such as it is the case with arginine, in correspondence with 15 atoms in the side chain of lysine)[8] is additionally enabled by the fact that fitting into the adequate quantum-level is necessary within both the columns and rows, as has already been said. But what is the most interesting here is the fact that the level has been achieved by „pure“ chemistry at the same time, as will be explained hereinafter.

---

[8] From the aspect of validity of the principle of self-similarity lysine which, apart from an $NH_2$ group, also has four $CH_2$ groups in the side chain, is more correspondent to the leucine (the first possible case of branched molecule) than are the derivatives with one, two, or three $CH_2$ groups.

## 4.2 The generation of the entire amino acid system

In order to give an explanation for just above presented question, let us take another look at Figure 1.2 and try to analyze the generation of the entire system of canonical AAs. The first possible amino acid must be Glycine (glycinic stereochemical type) with side chain of only one and that the smallest possible atom („copied“ from the „head“ of amino acid). But if that amino acid is an element of a system which is starting to generate itself, the following amino acid must be related to it and must be the first possible case of the zeroth cyclicity – Alanine, with $CH_3$ group as a side chain; more accurately with $CH_2$-**H** group, if $CH_2$ group is a characteristic of the subsystem (subsystem of alaninic stereochemical type) and if H represents the cognate „clasp“ both with the Glycine, and with the first possible substituent, with serine (**H** in $CH_2$-**H** group substituted by **OH** group). (With the generation of *oxygen* substituent – serine, in a „parallel“ process its *sulfurous* analogue, cysteine, is generated.)[9] The next two steps are: the first possible semi-cyclicity – valine (valinic stereochemical type: V & I) and the first possible full cyclicity – proline; proline as a prolinic stereochemical type (Rakočević and Jokić, 1996).

By generating the diastereoisomeric isoleucine with the side chain of a butyl group, leucine was inevitably generated with isobutyl group (which is actually, the first possible *branching*), as well as diastereoisomeric treonine ([H-C(OH)-CH3]) instead of the potential serine derivative ([H-C(OH)-H $\rightarrow$ H-C($CH_2$-OH)-H]). [Correspondent and parallel with the generation process of S-C, is the generation process of T-M.]

Upon the generation of serine, in the next step (parallel) there is a generation of aspartic acid (by „copying“ carboxylic group from the amino acid „head“) and its first possible derivative – glutamic acid; followed by the generation of amide derivatives of these two carboxylic amino acids (N & Q).

Along the other line, by generating leucine, phenylalanine is also generated since it possesses the same structural motif; phenylalanine, followed by its hydroxide derivative, tyrosine. [Phenylalanine possesses the structure of an isobutyl grupe, in other words it is generated from toluen and not from benzene, which is the reason that toluen's ring is the starting molecule for aromatic canonical amino acids.]

---

[9] All these chemical connections are naturally realized through bioprecursors, namely in the relations presented in three previous papers (Rakočević and Jokić, 1996; Rakočević, 1998a, 2004).

Hereby, all *invariant* AAs and their first possible derivatives and/or analogs – the *less variant* AAs – are generated. The remaining ones are only four AAs – two aliphatic and two aromatic ones, all four – the *more variant* AAs (Figure 5). Here we now return to the discussion on fitting arginine and lysine, so only two aromatic AAs remain which must be selected so that they fit into the "narrow passage", created by rows and columns in the quantum-level "71" and "81", respectively. [We should take note that here, when fitting into this „narrow passage", tryptophan follows the logic of generating phenylalanine (though only with one of its two rings), and the last amino acid, histidine, follows the logic of the second tryptophan ring (raising the degree of freedom further, exactly be how much pyrole differs from imidasole).]

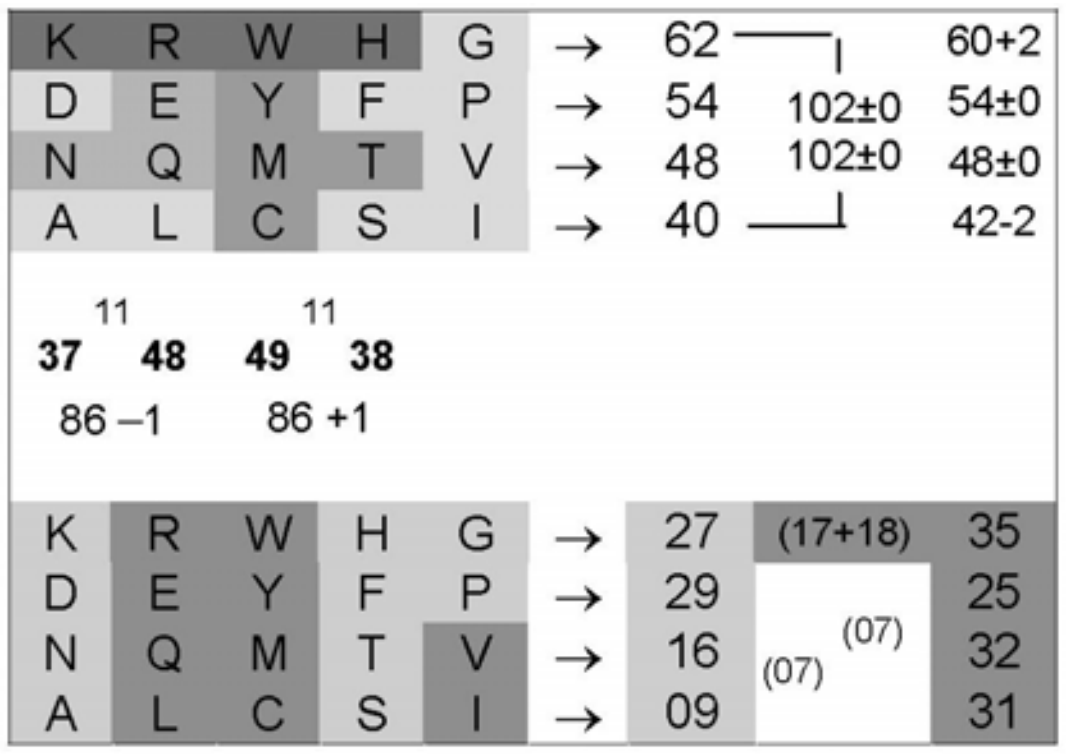

**Figure 5**. If we take as the "starting point" (above: dark-light as in Fig. 1.2) the "most variant" AAs, in the order which they have in the system of Gauss' algorithm (K,R,W,H) (Figure 1), and then if the remaining AAs are arranged according to chemical characteristics, a new arrangement is created. In this new arrangement the number of atoms in the columns is determined by Shcherbak's Prime quantum 37 (Shcherbak, 1994, 2003), while in the rows it is determined with the multiples of number 6. [Whereas the *number of atoms* is determined there by quantum 037, in the Shcherbak's system the *number of nucleons* is determined by that same quantum (in both amino acid classes, in four-codon and non-four-codon AAs) (Fig. 1 and Tab. 1 in Shcherbak, 1994).] Down: dark-light tones are given as in Figure 1; if so, then the determination through the minimum change principle is obvious (the change for a unit).

## 5 A NEW CLASSIFICATION OF AMINO ACIDS

Starting from analyzed relations in the amino acid system, presented in Figure 1.2, as well as from the finding that a „half" (10-1) of the set of canonical AAs is made as invariant, and the other „half" (10+1) as variant AAs, we come to a new classification of AAs, through a source splitting into „contact" and „non-contact" AAs (cf. Remark 1).

**Class 1**: *Invariant AAs* [**subclass 1**: contact AAs in two families; family 1: singlet AAs (G, P) and family 2: non-singlet AAs (V, I)[10]; **subclass 2**: non-contact AAs also in two families; family 1: aliphatic AAs (A, L, S, D) and family 2: aromatic amino acid (F); within family 1 there are two sub-families; sub-family 1: source aliphatic AAs (A, L) and sub-family 2: aliphatic derivatives (S, D)].

**Class 2**: *Variant AAs* [**subclass 1**: more variant AAs (family 1: aliphatic AAs K & R, and family 2: aromatic AAs W & H); **subclass 2**: less variant AAs (family 1: chalcogene AAs: Y & C and M & T; and family 2: carbonyl AAs: E, Q, N)]. (Cf. Figure 5 and Table 5.)

| | | | | | |
|---|---|---|---|---|---|
| (13) | **W, H** | 13 | 10 | Y, C | (8) |
| (5) | T, M | 12 | 15 | E, Q, N | (8) |
| (10) | *G, P,V,I* | *22* | *25* | *A, L, S, D , F* | (15) |
| | | | 20 | **K, R** | (8) |

**Table 5**. The determination by the number of carbon and hydrogen atoms. The number of atoms in the given sets of amino acid molecules (correspondent to the classification given in Section 5): carbon (in brackets) and hydrogen - within two columns, in the center of the Table (rows *b* and *c* in Table 7, respectively).

[10] Singlet amino acids – because a single amino acid is contained within one stereochemical type (Glycine in glicinic and Proline in prolinic type).

The Family of chalcogene AAs can be further classified; subfamily 1: OH and SH derivatives: Y & C, respectively; and subfamily 2: C-$CH_3$ and S-$CH_3$ derivatives: T & M, respectively.

The Family of carbonyl AAs also in two subfamilies classified: subfamily 1: non-nitrogen, E; and subfamily 2: nitrogen AAs, N & Q.

## 6 VARIABILITY IN RELATION TO POLARITY

Table 6 shows that above presented variability exists in a strict relation to polarity. In column with the designation "more", both invariant AAs (G & P) are semi-polar whereas four variant AAs appear to be in proportion 2:2; two aromatic AAs (W & H) are also semi-polar and two aliphatic (K & R) are polar. On the other hand, in column with the designation "less", we have a symmetric AAs distribution in proportion **5**:**2** / 2:5. Five invariant AAs as well as two variant (bold) are non-polar; in contrary, two invariant as well as five variant AAs (non-bold) are polar. It is important to say that four AAs (G, P & W, H), on the left side in Table 6, we take here as semi-polar through their ambivalence (*see* Footnote 3).

| more | less | |
|---|---|---|
| *G P* | **V I** | invariant |
| | **A L F** | |
| | S D | |
| *W H* | T N Y | variant |
| K R | E Q | |
| | **C M** | |

**Table 6**. Variability in relation to the polarity. Explanation in the text (Sections 5 and 6)

## 7 SPECIFIC ATOM NUMBER DETERMINATIONS

### 7.1 Determination with prime quantum 037

Once we know that out of 20 canonical AAs there are exactly four AAs with higher degree of freedom (more variant) than the remaining 16, then it is possible to organize the entire system in view thereof, as was shown in Figure 5. Thereby these four AAs should observe fundamental chemical hierarchy – the first ones are aliphatic (K, R), followed by aromatic (W, H) in the order we have in the system presented in Figure 1. Thereafter, there can be no more ambiguities. Two acidic AAs (D & E) follow two basic ones (K & R), followed by amide derivatives (N & Q) and the remaining two source aliphatic AAs (A & L). Yet, in continuation of two giving aromatic AAs (W, H), there are two remaining aromatic ones, two larger aliphatic chalcogenic ones (M, T) and finally two smaller chalcogenic ones (C, S). The position of contact AAs remains the same as in the system in Figure 1.

Atom number in this new system, as we see, is also[11] determinated by Shcherbak's prime quantum 037 (Shcherbak, 1994; 2003), with adequate other balances.

## 7.2 Determinations through atom number of carbon and hydrogen

A deeper analysis reveals the sense of the balance of number of carbon atoms in the sets of amino acidic molecules, given in Tab. 5. If we mark the numerical basis of a numbering system with the $q$, then the proportion $q/2 : q = 1: 2$ represents "the symmetry in the simplest case" (Marcus, 1989) for any $q$. However, only in the case when $q = 10$, therefore in the decimal numbering system, we also have the correspondence with the Golden mean as the best possible proportion and harmony. Namely, in the decimal numbering system, the $q/2$ is 5, and exactly through the square root of number 5 we come to two solutions of the square equation of the Golden mean. Even more, the „quantum 5" appears also within a „p-adic genetic information space" as an adequate „5-adic model for ... genetic code" (Dragovich and Dragovich, 2006).

Knowing this, it becomes understandable why number 5 is the starting quantum here (in the system presenting in Table 5). Starting from the number 5, there are two possible paths which stick to harmony. The first one is the realization of the following steps of Fibonacci sequence (which in itself is in correspondence with the Golden mean), achieving number 8 and number 13, which is the case here. The other one is the realization of multiples of number 5, for example, achieving number 10 and 15, which is also the case here.

---

[11] We say „also", because so far Shcherbak's "Prime quantum 037" was only known as a determinant of the number of nucleons in four-codon and non-four-codon AAs (Figure 1 in Shcherbak, 1994).

Certainly, the sense of the correspondence of the number of carbon atoms in a set of 20 canonical amino acids, with the numbers of Fibonacci sequence **5**, 8 and 13, becomes clear only when we know that the number of carbon atoms in two classes of that set (class I and class II as in Table 7) corresponds to the numbers of Fibonacci sequence 1, 2, 3 and **5** (cf. Table 2, last row, in Yang 2004, p. 1253).

| | | | | | | | | | | | |
|---|---|---|---|---|---|---|---|---|---|---|---|
| (i) | 117 | 131 | 121 | 147 | 146 | 149 | 131 | 204 | 181 | 174 | 1501 |
| (h) | 43 | 57 | 47 | 73 | 72 | 75 | 57 | 130 | 107 | 100 | 761 |
| **(g)** | **451** | **653** | **227** | **244** | **239** | **129** | **341** | **137** | **235** | **687** | **3343** |
| **(f)** | **432** | **631** | **213** | **225** | **219** | **109** | **319** | **110** | **211** | **661** | **03130** |
| **(e)** | **19** | **22** | **14** | **19** | **20** | **20** | **22** | **27** | **24** | **26** | **213** |
| (d) | 00 | 00 | 01 | 02 | 02 | 01 | 00 | 01 | 01 | 03 | (11) |
| (c) | 05 | 09 | 03 | 05 | 06 | 07 | 09 | 08 | 07 | 10 | 69 |
| (b) | 03 | 04 | 01 | 03 | 03 | 03 | 04 | 09 | 07 | 04 | 41 |
| **(a)** | **10** | **13** | **05** | **10** | **11** | **11** | **13** | **18** | **15** | **17** | **0123** |
| **I** | **V** | **L** | **C** | **E** | **Q** | **M** | **I** | **W** | **Y** | **R** | **I** |
| **II** | **G** | **A** | **S** | **D** | **N** | **T** | **P** | **H** | **F** | **K** | **II** |
| **(a)** | **01** | **04** | **05** | **07** | **08** | **08** | **08** | **11** | **14** | **15** | **81** |
| (b) | 00 | 01 | 01 | 02 | 02 | 02 | 03 | 04 | 07 | 04 | 26 |
| (c) | 01 | 03 | 03 | 03 | 04 | 05 | 05 | 05 | 07 | 10 | 46 |
| (d) | 00 | 00 | 01 | 02 | 02 | 01 | 00 | 02 | 00 | 01 | 09 |
| **(e)** | **10** | **13** | **14** | **16** | **17** | **17** | **17** | **20** | **23** | **24** | **171** |
| **(f)** | **448** | **436** | **639** | **219** | **217** | **432** | **424** | **213** | **205** | **223** | **3456** |
| **(g)** | **458** | **449** | **653** | **235** | **234** | **449** | **441** | **233** | **228** | **247** | **3627** |
| (h) | 01 | 15 | 31 | 59 | 58 | 45 | 41 | 81 | 91 | 72 | 494 |
| (i) | 75 | 89 | 105 | 133 | 132 | 119 | 115 | 155 | 165 | 146 | **01234** |

**Table 7**. The most important parameters for 20 canonical amino acids. I. The first class of amino acids, handled by the first class of enzymes aminoacyl-tRNA synthetases; II. The second class of amino acids, handled by the second class of enzymes aminoacyl-tRNA synthetases. Both classes in the AAs pairs as in Rakočević, 1998b (Survey 4); the order is dictated by the number of atoms in the side chain of each first member of an amino acid pair; (a) number of all atoms in the "bodies" i.e. side chains of amino acid molecules; (b) number of carbon atoms in the "bodies", i.e. side chains of amino acid molecules; (c) number of hydrogen atoms in the "bodies" i.e. side chains of amino acid molecules; (d) number of non-hydrocarbon atoms in the "bodies", i.e. in the side chains of amino acid molecules; (e) number of all atoms in the whole amino acid molecules ("head" plus "body"); (f) number of all atoms in the amino acid belonging codons; g = e + f; (h) number of nucleons in the "bodies", i.e. side chains of amino acid molecules; (i) number of nucleons in the whole amino acid molecules ("head" plus "body").

The number of hydrogen atoms (two columns of numbers in the central part of Table 5) is also determined by the multiples of number 5 and only by the final

Fibonacci's number (final in the Fibonacci sequence 1-13), as well as the numbers which are created with the unit change in the first and/or second position of that number (numbers 12 and 22 as 13 – 01 and 12+10, respectively).

## 7.3 Determinations through atom number of non-hydrocarbon elements

Apart from the determinations with the atom number of carbon and especially the atom number of hydrogen, the determination with atom number of non-hydrocarbon elements appear to be exact and relevant too. Namely, the system of amino acid doublets such as we find in Table 1.1, organized in the system of quartets (pairs of pairs), by preserving original hierarchy (Figure 1), “hides” in itself not only the Gauss’ algorithm, but another specific determination which essentially comes down to determination by the sequence of even numbers 0-2-4-6-8 (Figure 6 in relation to row *d* in Table 7): 2 & 4 in outer and 6 & 8 in inner rows (10 in odd and 10 in even rows). [*Remark 3*: As a noteworthy is the fact that the sum of the squares of the number of atoms within side chains of 20 amino acid molecules is determined also with the sequence 0-2-4-6-8; equals exactly 02468: G(01) + A(16) + S(25) + C(25) + D(49) + N(64) + T(64) + P(64) + V(100) + E(100) + Q(121) + H(121) + M(121) + L(169) + I(169) + F(196) + Y(225) + K(225) + R(289) + W(324) = 02468.]

| | | (a) | | | 10 | | 08 | 11 | (b) | | 09 |
|---|---|---|---|---|---|---|---|---|---|---|---|
| S01 | T01 | L00 | A00 | G00 | **02** | (02) | 04 | 01S | 01T | 01M | 01C |
| D02 | E02 | M01 | C01 | P00 | **06** | (02) | 08 | 02D | 02E | 02Q | 02N |
| K01 | R03 | Q02 | N02 | V00 | **08** | (04) | 04 | 01K | 03R | 00L | 00A |
| F00 | Y01 | W01 | H02 | I00 | **04** | (00) | 04 | 00F | 01Y | 01W | 02H |
| 04 | 07 | 04 | 05 | 00 | | | | 00G | 00V | 00I | 00P |
| **10+1** | | **10-1** | | | 10 | | 12 | 00 | | 00 | |
| | | | | | | | | (10+1) | | (10-1) | |
| (light tones: **10-1**; dark tones: **10+1**) | | | | | | | | | | | |

**Figure 6**. Relations of variability and polarity expressed through the number of non-hydrocarbon atoms. Everything is the same as in Figure 1, the only difference is that in the former case all atoms within amino acid side chains are given, and in the latter case only the atoms which are neither the atoms of carbon nor the ones of hydrogen.

## 8 DETERMINATION WITH THE SEQUENCE OF NATURAL NUMBERS

We can understand the deeper sense of determination with the sequence 0-2-4-6-8, i.e. sequence of even numbers, only when we notice that other sequences generated from the sequence of natural numbers are also in the "game". [By this one must bear in the mind that in GCT not only codons but also amino acids possess a strict order in correspondence to natural numbers sequence, from 0 to 19 (Damjanović, 1998; Damjanović & Rakočević, 2005).] So, *the atom number* in class I of AAs, in side chains of 10 AAs, is determined by the sequence "0123" (row *a* in Table 7, above). On the other side, *the nucleon number* in class II of AAs in whole molecules of 10 AAs, is determined by the sequence "01234" (row *i* in Table 7, down). (Class I and class II of AAs in correspondence to two classes of enzymes aminoacyl-tRNA synthetases.)

The number of atoms within the 32 codons coding for AAs in class II equals 3456 (row *f* in Table 7, down). [The difference 3456 – 1234 = 2222 shows that the number of *atoms* within the codons is greater for 2222 than the number of *nucleons* within corresponding amino acid molecules.] Except that, there is the same number of atoms (3456) in codon nucleotides, within two and two columns of GCT (Figure 7, down). Further, atom number within side chains of 23 amino acid molecules, in a symmetric arrangement through rows and columns in GCT, is determined by the sequence "456/789" (Figure 7, on the right side; *see* also these relations in Verkhovod, 1994, Figure 2).

All these insights come down to the fact that we must comprehend (and understand) that in the genetic code, apart from two known alphabets (4 amino basic and 20 amino acidic molecules), there is also another one, the original and fundamental (hidden?) alphabet – the sequence of natural numbers. What else could be expected if we know that all molecules are made of atoms of chemical elements (within Periodic system) whose structural hierarchy is harmonized with that very sequence of natural numbers.

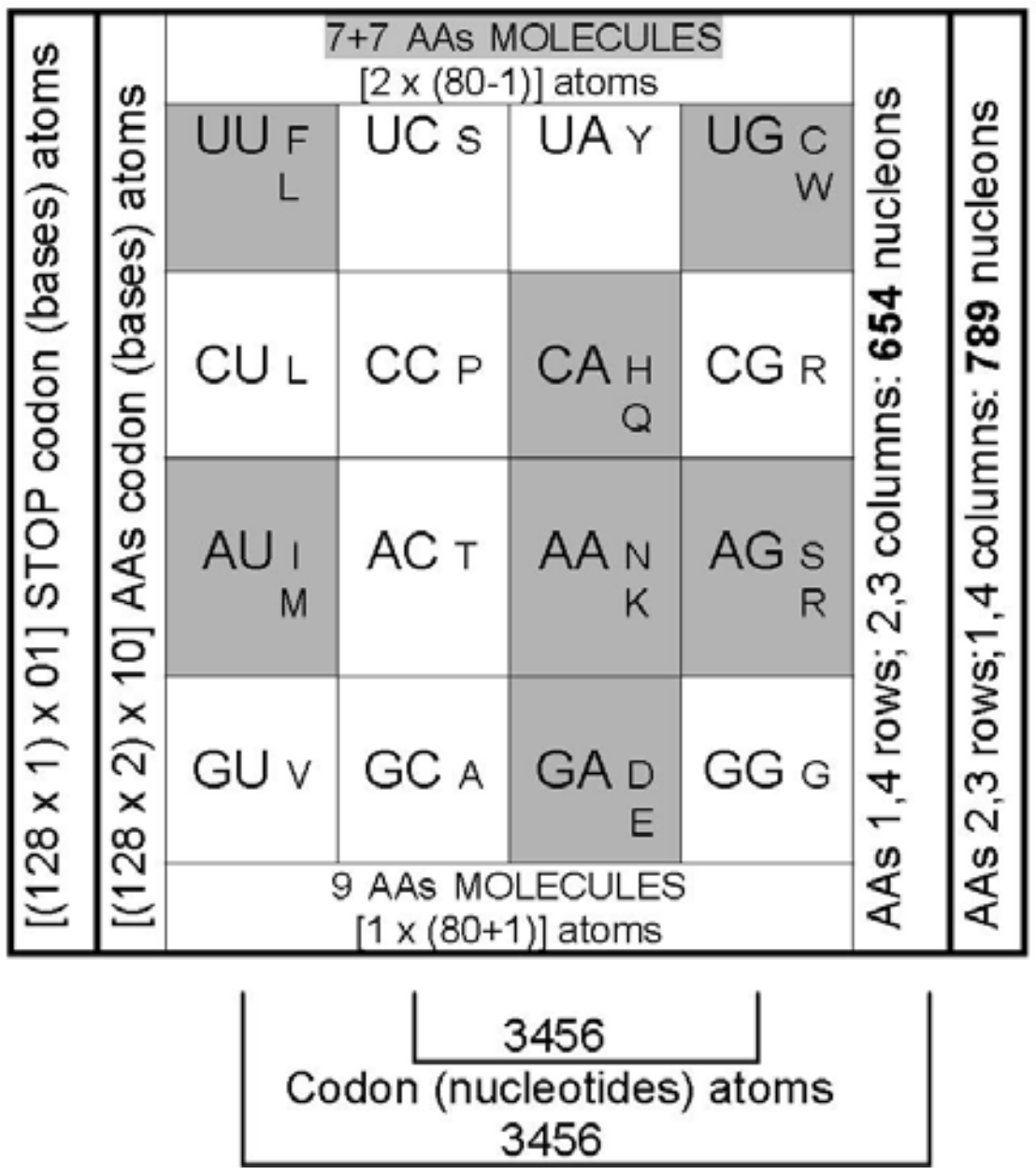

**Figure 7.** The number of atoms and nucleons. In the GCT the codon doublets code for one or two amino acids. The number of atoms and nucleons in the classes and subclasses of amino acids corresponds to the parts of sequences of natural numbers, with principle of minimum (unit) change. The external vertical row on the left: the number of atoms in three "stop" codons, calculated to bases (U = 12; C = 13; A = 15; G = 16). The internal vertical row on the left: number of atoms in the remaining 61 codons, which have amino acid meaning. Two vertical rows on the right: the number of nucleons in the GCT according to Verkhovod, 1994, Figure 2. Dark tones: double meaning codon doublets, each with two encoded amino acids (total: two times 8-1). Light tones: single meaning codon doublets, each with one encoded amino acid (total: 8+1 molecules, with 81 atoms as in class II, in row *a* of Table 7). The bottom of Figure shows that in two and two columns, i.e. in two and two rows, there are 3456 atoms in the codons, calculated to nucleotides (UMP = 34; CMP = 35; AMP = 37; GMP = 38).

## 9 CONCLUSION

From the presented facts and given discussion in previous eight sections follows that it make sense to speak about genetic code as a harmonic system. On the other side, presented harmonic structures which appear as the unity and coherence of *form* (atom and nucleon number balances) and *essence* (physicochemical properties) provide evidence to support the hypothesis, given in a previous paper (Rakočević, 2004), that genetic code was complet**e** from the very beginning as the condition for origin and evolution of the life.

## REFERENCES

Damjanović, Z., 1998. Logic core of genetic code. Glasnik Sect. Nat. Sci. Montenegrin Acad. Sci. Art (CANU) 12, 5-8.

Damjanović, M. Z., Rakočević, M. M., 2005. Genetic code: an alternative model of translation. Ann. N. Y. Acad. Sci. 1048, 517-523.

Doolittle, R.F., 1985. Proteins. Scientific American, 253, 74-85.

Dragovich, B., Dragovich, A., 2006. A p-Adic Model of DNA Sequence and Genetic Code. q-bio. GN/0607018.

Falconer, K., 1990. Fractal geometry. John Wiley, New York pp. 34-52.

Jukes, T. H., 1973. Possibilities for the evolution of the genetic code from a preceding Form. Nature 246, 22-27.

Konopel'chenko, B.G., Rumer, Yu. B., 1975. Klassifikaciya kodonov v geneticheskom kode. Dokl. Akad. Nauk. SSSR. 223, 471-474.

Kyte, J., Doollittle, R. F., 1982. A simple method for displaying the hydropathic character of a protein. J. Mol. Biol. 157, 105-132.

Marcus, S., 1989. Symmetry in the simplest case: the real line. Comput. Math. Appl. 17, 103-115.

Moore, G.A., 1994. The limit of the golden numbers is 3/2. The Fibonacci Quaterly, June-July, 211-217.

Popov, E. M., 1989. Strukturnaya organizaciya belkov. Nauka, Moscow (in Russian).

Rakočević, M. M., 1997. Genetic code as a unique system. SKC, Niš pp. 60-64.

Rakočević, M. M., 1998a. Whole-number relations between protein amino acids and their biosynthetic precursors. J. Theor. Biol. 191, 463 – 465.

Rakočević, M. M., 1998b. The genetic code as a Golden mean determined system. Biosystems 46, 283-291.

Rakočević, M. M., 2004. A harmonic structure of the genetic code. J. Theor. Biol. 229, 463 – 465.

Rakočević, M. M., Jokić, A., 1996. Four stereochemical types of protein amino acids: synchronic determination with chemical characteristics, atom and nucleon number. J. Theor. Biol. 183, 345 – 349.

Shcherbak, V. I., 1994. Sixty-four triplets and 20 canonical amino acids of the genetic code: the arithmetical regularities. Part II. J Theor. Biol. 166, 475-477.

Shcherbak, V. I., 2003. Arithmetic inside the universal genetic code. Biosystems 70, 187-209.

Swanson, R., 1984. A unifying concept for the amino acid code. Bull. Math. Biol. 46, 187-207.

Verkhovod, A. B., 1994. Alphanumerical divisions of the universal genetic code: new divisions reveal new balances. J. Theor. Biol. 170, 327-330.

Woese, C.R., et al., 1966. On the fundamental nature and evolution of the genetic code. In: Cold Spring Harbor Symp. Quant. Biol., 31, 723-736.

Yang, C.M., 2004. On the 28-gon symmetry inherent in the genetic code intertwined with aminoacyl-tRNA synthetases – The Lucas series. Bull. Math. Biol. 66, 1241-1257.